\documentclass[article ,11pt]{amsart}
\usepackage{amsmath}
\usepackage{amssymb}
\usepackage[foot]{amsaddr}
\usepackage[usenames,dvipsnames]{xcolor}
\usepackage{subfigure}
\usepackage{lineno}
\usepackage[paperwidth=7in,paperheight=10in,text={5in,8in},left=1in,top=1in,headheight=0.25in,headsep=0.4in,footskip=0.4in]{geometry}

\definecolor{linenocolor}{gray}{0.6}
\definecolor{prec}{RGB}{42,115,205}
\definecolor{sens}{RGB}{255,112,0}
\definecolor{spec}{gray}{0.4}

\usepackage{fix-cm}

\usepackage{url}
 \usepackage{multirow}
\usepackage[numbers, sort&compress]{natbib}
\usepackage{graphicx}
\usepackage{dcolumn}
\usepackage{bm}

\begin{document}


\title[Social Conformity Despite Preferences for Distinctiveness]{Social Conformity Despite Individual Preferences for Distinctiveness}

\author[Smaldino \& Epstein]{Paul E. Smaldino$^{1,2}$ \and Joshua M. Epstein$^1$}
\email{paul.smaldino@gmail.com, jepste15@jhmi.edu}
\address{$^1$ Center for Advanced Modeling in the Social, Behavioral, and Health Sciences, Johns Hopkins University, 5801 Smith Ave, Baltimore, MD 21209}
\address{$^2$ Department of Anthropology, University of California, Davis, 1 Shields Ave, Davis, CA 95616}


\begin{abstract}
We demonstrate that individual behaviors directed at the attainment of distinctiveness can in fact produce complete social conformity.  We thus offer an unexpected generative mechanism for this central social phenomenon. Specifically, we establish that agents who have fixed needs to be distinct and adapt their positions to achieve distinctiveness goals, can nevertheless self-organize to a limiting state of absolute conformity. This seemingly paradoxical result is deduced formally from a small number of natural assumptions, and is then explored at length computationally. Interesting departures from this conformity equilibrium are also possible, including divergence in positions. The effect of extremist minorities on these dynamics is discussed. A simple extension is then introduced, which allows the model to generate and maintain social diversity, including multimodal distinctiveness distributions. The paper contributes formal definitions, analytical deductions, and counterintuitive findings to the literature on individual distinctiveness and social conformity.

\end{abstract}


\maketitle


\linenumbers
\modulolinenumbers[2]

\section{\label{sec:level1}Introduction}

Few would dispute that we humans make appraisals of our individual ``distinctiveness," that we differ in our needs to appear distinct, and that we take actions (e.g., we alter our appearance or expressed opinions) to attain our distinctiveness goals. Furthermore, preferences for distinctiveness and the associated remedial adaptation strategies are at work in the formation of social groups and networks, and in other cultural dynamics such as assimilation or polarization. However, in the broad literature on the psychology and sociology of distinctiveness \cite{berg08, simm57, snyd80, brew91, lynn97, kim99, horn04, jett04, timm08, imhoff09, vig11}, there is little mathematical precision in defining ``distinctiveness preferences," and little explicit modeling of the individual behaviors adopted to satisfy them or the collective dynamics generated by these individual adaptations. 

Most formal models dealing with individual preferences for differentiation posit strict anti-conformity, in which agents adopt whatever position constitutes the minority at a given time \cite{galam04, lama05, nyc13, mas10, jav2014}. Such individual behavior of course endogenously alters the distribution of positions and can produce interesting social dynamics. But it precludes the emergence of conformity, our core concern. Relatedly, Smaldino et al. \cite{smaldino12} modeled individuals with preferences for membership in groups with different degrees of numerical predominance but this model was not concerned with individual differences or distinctiveness {\em within} a population. 

To our knowledge, no previous model has formally defined distinctiveness preferences as we do, or shown that the pursuit of distinctiveness, thus defined, can lead to conformity. We provide simple and intuitive formal definitions of distinctiveness preferences and the individual actions aimed at satisfying them, and show that these yield counterintuitive social dynamics. Foremost among these is that a population of agents with {\em fixed} needs to be distinct can self-organize into a state of strict conformity. We then show how an elementary, but also novel, extension of the model facilitates sustained diversity in attributes of interest. We conclude with a brief discussion of several lines for future research.

\section{\label{sec:level1}Model 1: Distinctiveness in Units of Standard Deviation}

To keep the exposition as simple as possible, we imagine a fixed population of $N$ agents.  They have only the following four very simple attributes: a position, information about the distribution of positions, a distinctiveness goal, and an adjustment rule. Time in this model is discrete. 

\begin{quote}
(1) {\bf Position:} At every time, $t$, each agent has an observable ``position," $x_i(t)$. Position is a one-dimensional real-valued feature that agents can adjust.  
\end{quote}

For example, position could be an expressed taste or location on a left-right political spectrum. It is not a location in any landscape, network, or other physical coordinate system.  This version of the model is spaceless in that sense. Agents interact only with aggregate variables -- the mean and standard deviation of a distribution. They do not interact directly with one another. One could of course introduce spatial coordinates, neighborhoods, networks, and direct interaction with other agents, but we hold off on these complexities here. 

\begin{quote}
(2) {\bf Information:} Each agent is assumed to know its own position, and to correctly estimate (i.e., to intuit) the mean $\bar{x}(t)$ and standard deviation $\sigma(t)$ of positions in the population.  
\end{quote}

Given a distribution of positions, it strikes us as natural to define distinctiveness in terms of deviations from the mean. The simple intuition is that, in a drab office where jet black suits are the norm, a dark gray one may turn heads, whereas the ambient diversity of Times Square requires far more flamboyance to be noticed. So, a fixed distinctiveness preference (in standard deviations) may elicit radically different behaviors in different settings. Accordingly, we define ideal distinctiveness as follows:

\begin{quote}
(3) {\bf Ideal Position:} Agents have a fixed and unobservable distinctiveness preference, $\delta_i$. This parameter characterizes the individual's ideal position; it is not absolute, but relative to other agents in the population. Specifically, at any time, the $i$th agent's ideal position $x_i^*(t)$ is given by
\begin{equation}
x_i^* (t) = \bar{x}(t) + \delta_i \sigma(t),
\end{equation}
where $\sigma(t)$ is the standard deviation of the population's current positions.
\end{quote}

A positive value for $\delta_i$ indicates an ideal position $\delta_i$ deviations above (e.g., to the political right of) the population mean; a negative value indicates an ideal position $\delta_i$ deviations below it.  Some people might need to be three sigmas ($\delta_i = 3$) from the mean; others are content to hover near the average. We do not assume that any individual is consciously aware of, or could ``tell you," their delta, only that these preferences exist, and that agents adapt to satisfy them.

Note that, because ideal position is by definition relative to the distribution of other agentsÕ positions, different positions may be satisfactory (i.e., equal to an agent's ideal position) at different times, corresponding to different distributions of positions in the population. By the same token, a specific position may confer ideal distinctiveness today, but not tomorrow, if the positions of other agents shift. Of course, individuals may find themselves in positions that fail to satisfy their need for distinctiveness. In such cases, they adjust their positions to better satisfy this need. As the simplest (error-correction type) mechanism, we posit the following

\begin{quote}
(4) {\bf Positional Adjustment Rule:} At time $t + 1$, individuals adjust their positions at a rate proportional to their distance from the ideal at time $t$.  Specifically, with $x_i^* (t)$ given by Eq. 1, each individual updates her position according to
\begin{equation}
x_i (t+1) = x_i (t) + k [x_i^* (t) - x_i (t)]
\end{equation}
where $0 < k < 1$ is an adjustment rate. (We exclude 0 and 1 since they, respectively, cancel all dynamics or impose equilibrium in one step).
\end{quote}

So, the farther is your current position from that which would satisfy your (fixed) need for distinctiveness, the greater is your adjustment in observable position. Equivalently, under this rule, an individual far from her ideal position will move faster than an individual close to it. Several extensions and refinements are discussed later, but this (1)--(4) is the complete agent specification. Though spare, the range of social dynamics is surprisingly rich. 

\subsection{Behavior of the Mean}
The first question one might pose is: Given a population of individuals with initial positions $x_i(0)$ and distinctiveness preferences $\delta_i$, how will the mean position behave? After a modicum of algebra (see Appendix B), we derive the change in the mean position $\Delta \bar{x}$ to be given by
\begin{equation}
\Delta \bar{x} = \frac{k}{N} \sum_i [\bar{x} (t) + \delta_i \sigma(t) - x_i (t)],
\end{equation}
which, by Eq. 1, is simply 
\begin{equation}
\Delta \bar{x} =  \frac{k}{N} \sum_i [x_i^* (t) - x_i (t)].
\end{equation}
Eq. 4 implies that the mean position will not change if each individual is in an optimally distinct position, because if $x_i = x_i^*$, the right hand side is zero. A different type of equilibrium is where agents are all placed at the {\em same} position. Whatever that common position may be, the mean will not change because in this case the mean is the common position and the standard deviation is zero. So, by Eq. 3, there is no change in position. In this sense, any possible position is an equilibrium, and the common position in which they are placed will be regarded by all agents as ideal. However, so long as initial positions are not ideal or identical, no equilibrium is strictly attained, because the variance will never reach zero, but only approach it as a limit. Our first and central result, whose robustness we explore in a number of settings, is that convergence toward this conformity limit occurs despite positive distinctiveness preferences. 

\begin{figure*}[tp]
\includegraphics[width=0.99\textwidth]{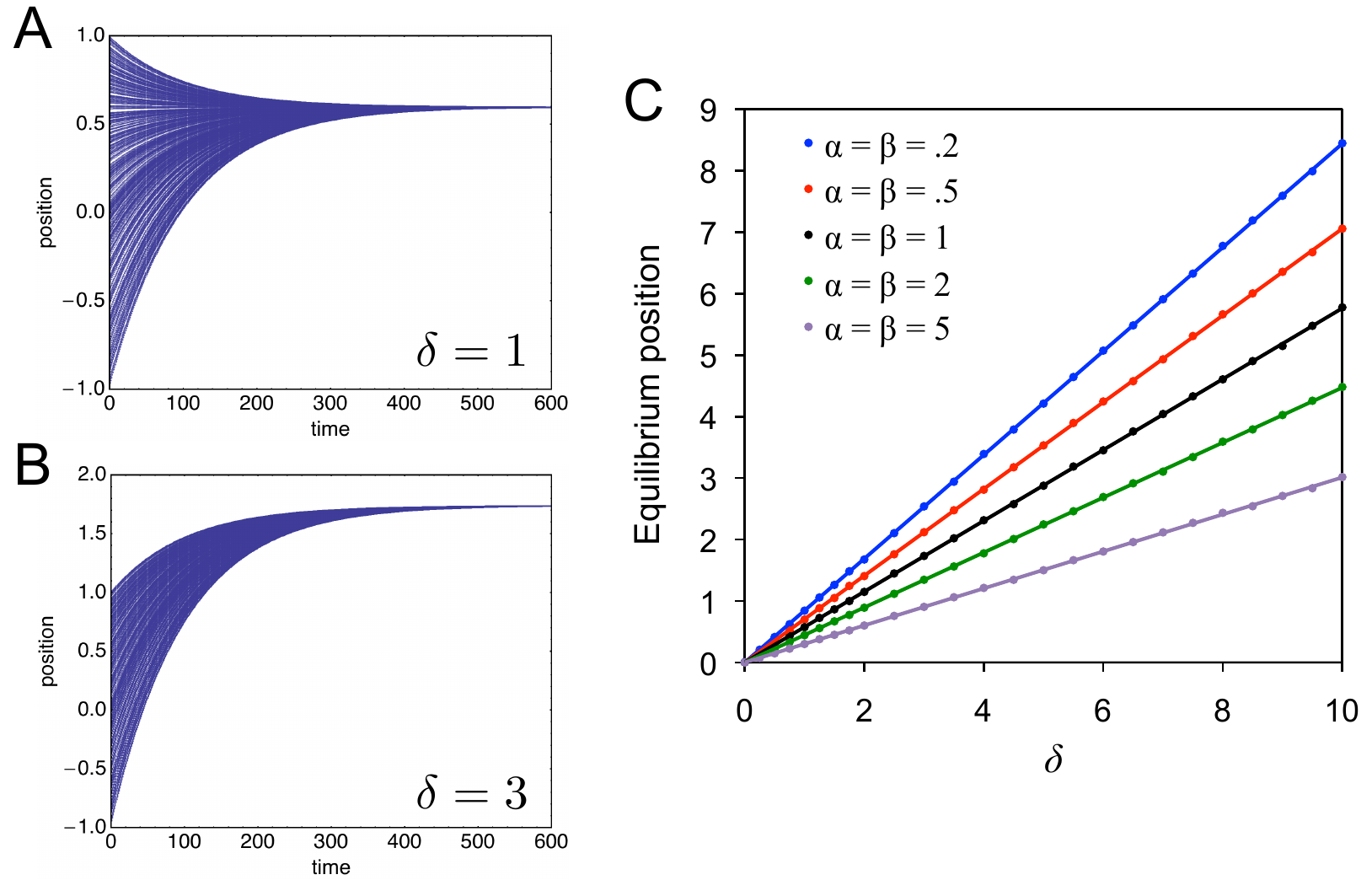}
\caption{\label{fig:fig1} The system converges on a single value when all individuals share the same preferred distinctiveness. We first illustrate example trajectories for (A) $\delta = 1$ and (B) $\delta = 3$. (C) Equilibrium positions at convergence limit as a function of $\delta$ for several distributions of initial positions. For these and all other runs, $N = 500$, $k = 0.01$.}
\end{figure*}

\subsection{Conformity Despite A Single Global Preference for Distinctiveness }
We begin with the simplest case, in which all individuals have the same fixed positive distinctiveness preference, $\delta$ (i.e., $\forall i, \delta_i = \delta > 0$). Eq. 3 then becomes simply:
\begin{equation}
\Delta \bar{x} = k \delta \sigma (t).
\end{equation}
Since $k \delta$ is a constant, the rate of change in the mean position will be proportional to the standard deviation of individual positions. But, what becomes of the standard deviation itself? It can be proven (see Appendix C) that
\begin{equation}
\sigma^2 (t+1) = (1-k)^2 \sigma^2 (t). 
\end{equation}
This first-order difference equation is solvable analytically for the time evolution of variance. Starting with any specified initial variance $\sigma^2 (0)$, we have
\begin{equation}
\sigma^2 (t) = (1-k)^{2t} \sigma^2 (0). 
\end{equation}

Eq. 7 makes clear that conformity (the zero variance state) is a limit. As noted earlier, if all agents are placed at a point, no one will depart. So, any position can be an equilibrium. But because $k$ is a real number strictly less than one, these equilibria, though attractors, are not strictly attainable from non-equilibrium positions\footnote{However, it can  be shown that, when $k$ is small and $\delta$'s are identical, the equilibrium position is well approximated by $\delta \sigma(0)$.}. For any specified real number $z > 0$, however small, there is a time $t_z$ after which the standard deviation is less than $z$.  In any particular case, the waiting time will depend on $k$, $\delta$, and other factors. As it does not affect our thrust here, waiting time will not be further pursued.

\begin{figure*}[tp]
\includegraphics[width=0.99\textwidth]{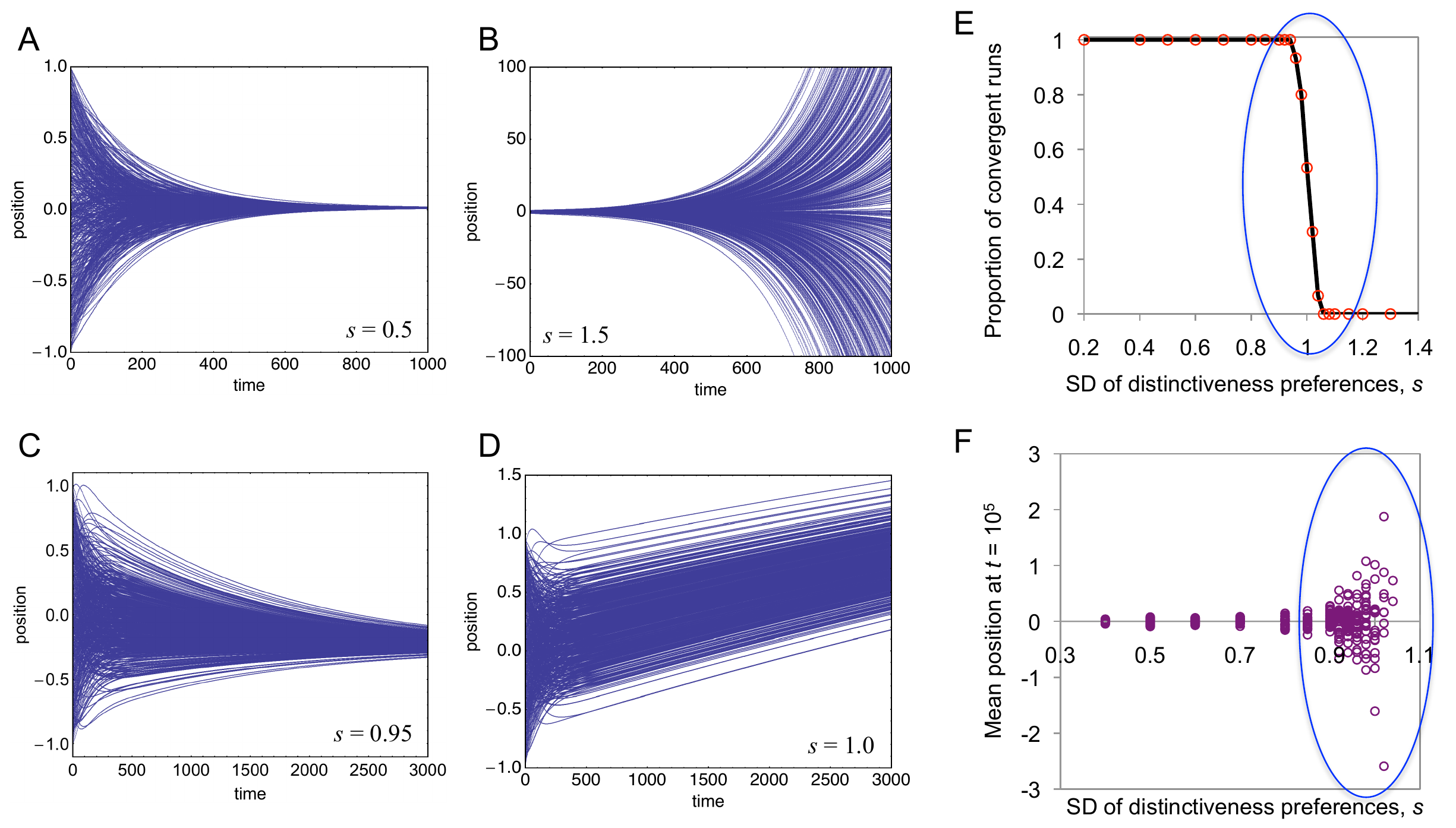}
\caption{\label{fig:fig2} Individual preferences drawn from a normal distribution. (A--D) Example trajectories of runs, with $s$ as indicated. (E) The proportion of simulation runs (of 30) for which the population converged as a function of $s$. (F) Mean population position for convergent runs as a function of $s$. This figure omits one outlier at (.96, $-5.46$). }
\end{figure*}

Notice that, although by Eq. 5 the mean position depends on $\delta$, by Eq. 7 the variance does not, but rather changes at a rate determined entirely by $k$. This difference between the two moments' dynamics is illustrated in Figs 1A and 1B. In both cases, initial positions are randomly drawn from the uniform distribution U($-1$,1). The variance converges to zero in both cases, but the mean increases more with $\delta = 3$ than it does with $\delta = 1$, as the above analytics would predict. Fig. 1C illustrates that the exact limiting position depends on both $\delta$ and the initial distribution of positions. Here, initial positions are drawn from various Beta distributions, transformed so that the support is [$-1$, 1] (The uniform distribution is recovered when $\alpha = \beta = 1$; see Fig. A.1). For each such distribution we plot the single common position toward which all agents converge as $\delta$ is increased from zero to 10. {\em Despite fixed common preferences for individual distinctiveness, the population approaches global conformity in position.}

Famously, Schelling's \cite{schel71} model showed that a macroscopic pattern of segregation does not warrant the inference that all individuals are discriminatory. Here, a macroscopic pattern of conformity does not mean that individuals lack desire for distinctiveness. Indeed, even if all agents have the same positive fixed preference for distinctiveness, and adjust their positions in proportion to their distance from this ideal, complete social conformity occurs and persists (it is approached monotonically as a limit). Fig. 1C also indicates that, given a perturbation from any equilibrium position, variance will again collapse to zero, but with agents converging toward a new common position. The model thus offers an unexpected generative mechanism for conformity. We now explore its robustness to selected variations.

\subsection{Heterogeneous Preferences for Distinctiveness Produce Bifurcation in Dynamics}

In the scenarios presented thus far, agents had the same positive distinctiveness preference, $\delta$. A perhaps more plausible presumption is that individuals vary in their preferences for distinctiveness. Will this change our results? We begin exploring this question by assuming distinctiveness preferences to be normally distributed with a mean of zero and a standard deviation of $s$. Dynamics are not as tractable analytically as in the previous case, and we turn to agent-based simulations. For this and all subsequent simulations, initial positions are drawn from U($-1$,1), $N = 500$, and the adjustment rate $k$ = 0.01. Before, with all agents having the same $\delta$, we saw convergence to conformity. Is this result robust to heterogeneous $\delta$s?

Strikingly, simulations indicate that to some extent, it is! Even with heterogeneous preferences for distinctiveness, the population can still converge to a single position (Fig. 2A). However, if the distribution of preferences is wide enough, the population instead diverges, with individual positions growing ever farther apart (Fig. 2B). Thus, as we increase this heterogeneity, a type of bifurcation occurs. For the normal distribution of preferences used in these simulations, this bifurcation occurs around $s = 1$. Specifically, for $s < 1$, positions converge. For $s > 1$, positions diverge. For values of $s$ very close to 1, we find that some runs converge and some diverge, due to noise in the particular distributions of initial positions and distinctiveness preferences (Fig. 2E).

The speed of convergence or divergence was slowest near this tipping point. This illustrates an interesting feature of the model. When the standard deviation of distinctiveness preferences is near the critical value of one, the population mean can change rapidly as individuals (asynchronously) update their positions, while their positions {\em relative to the mean} will change very slowly. Fig. 2D illustrates that although convergence is assured, extremely long periods of ``quasi-stability" can be maintained, during which the mean increases. We examined the limiting point of population convergence for heterogeneously distributed $\delta$s for runs in which the population converges (i.e., runs in which the standard deviation of positions was continuously decreasing after an initial transient period of reorganization, which generally lasted about 150 time steps). We found that for convergent runs, the closer was $s$ to the critical threshold, the farther from the initial population mean was the point of convergence proper, as seen in Fig. 2F. 
	
These computational findings are supported by analytical results with two agents, having distinctiveness preferences equal in magnitude but opposite in sign (i.e., $\delta_1 = -\delta_2$). For this case, it can be proven that two agents will converge if and only if the standard deviation of the agents' positions is less than one (see Appendix E).

\begin{figure}[tp]
\includegraphics[width=0.69\textwidth]{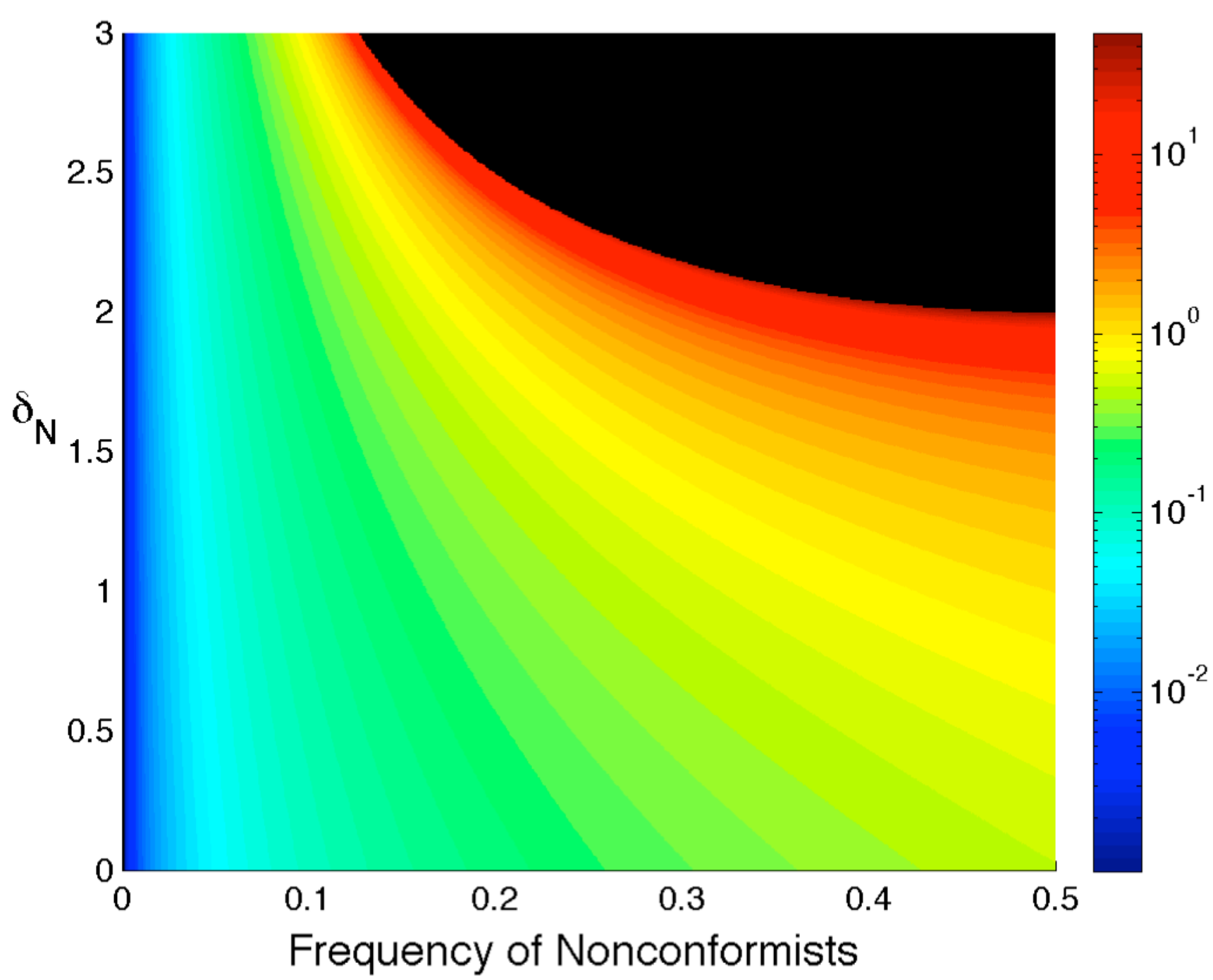}
\caption{\label{fig:fig3} Convergence positions for a minority of nonconformists. Initial positions were zero for conformists and 1 for nonconformists. The color map is on a log scale and indicates the limiting equilibrium position. That is, the color indicates the point toward which the conformists and nonconformists converge. The black area indicates values for which the system diverges. }
\end{figure}

\begin{figure*}[tp]
\includegraphics[width=0.99\textwidth]{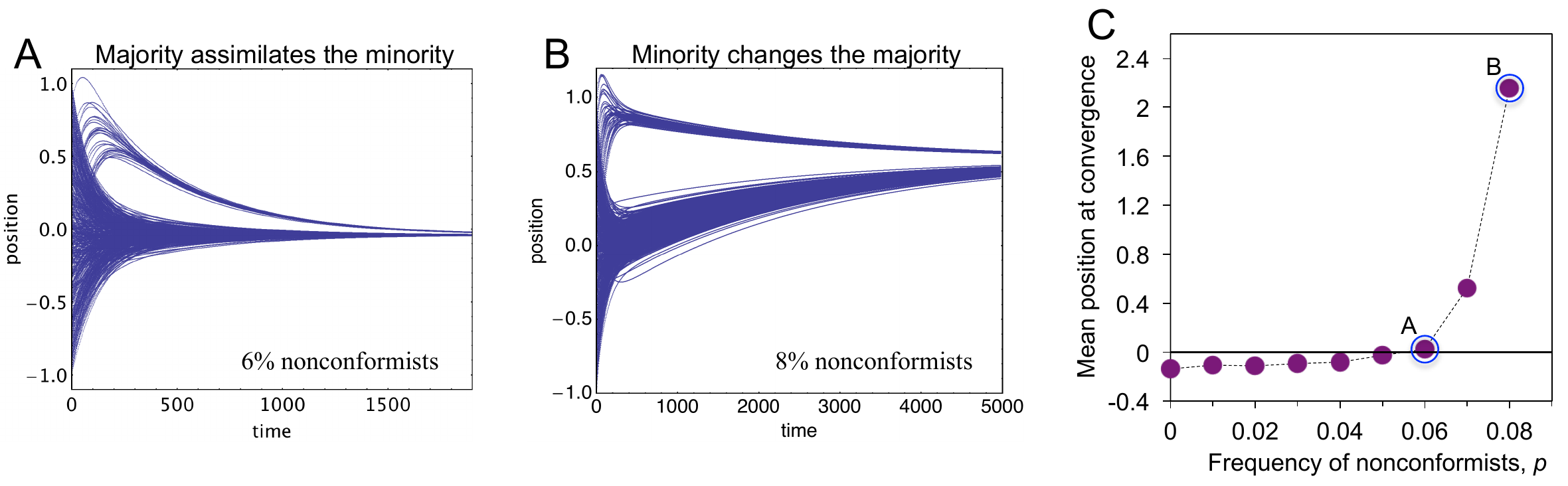}
\caption{\label{fig:fig4} A majority of conformists and a minority of extreme nonconformists. (A-B) Two example trajectory plots. (C) The mean position at convergence across 30 simulation runs as a function of the size of the majority group. The population rarely converged with more than 8\% nonconformists.}
\end{figure*}

\subsection{Bimodal Distributions of $\delta$s: A Minority of Nonconformists}

In the previous section we assumed a unimodal distribution of $\delta$s. However, it is also possible that individuals would cluster around different preferred degrees of distinctiveness. There are many possible multimodal distributions; we will consider two cases here. 

We first treat the simplest case of a population with exactly two types of individuals: conformists and nonconformists. Conformists prefer to be at the population mean ($\delta_C = 0$). Nonconformists prefer to be distinct ($\delta_N > 0$), but are otherwise identical. This case can be handled analytically. There are two free parameters: $\delta_N$, and the frequency of nonconformists in the population, $p$. In order to get dynamics, we assume that conformists and nonconformists start at different initial positions, so that the standard deviation of positions is nonzero. 

For a wide range of conditions, the system again converges toward a single value (global conformity). Fig. 3 shows that the limiting position will be increasingly far from the initial position of the conformists when there are more nonconformists (larger $p$) or when nonconformists have more extreme preferences for distinctiveness (larger $\delta_N$). Moreover, for every value of $p$, there is a critical value of $\delta_N$, above which the population does not converge. Specifically, this critical value is given by
\begin{equation}
\delta_N^* = \frac{1}{\sqrt{p (1-p)}}
\end{equation}
(see Appendix D for derivation). Above this critical curve, the conformists and nonconformists will grow ever farther apart, and their absolute positions will continue to increase. The area including and above this separatrix is colored black in Fig. 3.

Continuing to elaborate the model, we now explore dynamics assuming heterogeneous distinctiveness preferences not with just two values as above, but with values clustered about two distinct modes. For convergent runs, two possibilities suggest themselves. First, the majority, with preferences for very moderate distinctiveness (conformists) could assimilate the minority of individuals, who have more extreme preferences for distinctiveness (nonconformists). Second, a minority of nonconformists, though small, could dramatically influence the positions of conformists, moving them far from their initial positions. To explore these possibilities, we ran simulations in which each agent's distinctiveness preference was drawn from one of two possible continuous distributions. Nonconformists' $\delta$s were drawn from a normal distribution with a mean of 3 and a standard deviation of 0.1. In order to highlight the pull of nonconformists, we let the mean $\delta$ of the conformists be negative. Convergence to a positively valued position would thus indicate a strong influence of nonconformists. Conformists' $\delta$s were drawn from a normal distribution with a mean of $-0.2$ and a standard deviation of 0.3.  Obviously, these numerical choices are purely illustrative.

In this computational experiment, when nonconformists were a small minority, they tended to be assimilated into the majority as the population converged toward conformity, as in Fig. 4A. However, a slightly larger minority of nonconformists, still only 8\% of the population, exerted a much larger influence on the majority (Fig. 4B). Fig. 4C shows the limiting position of the population (for runs that converged) for differently sized minority groups. This illustrates another central finding: {\em A small minority of extreme nonconformists can exert large influences on a population}.

\section{\label{sec:level1}Model 2: Preferences for Absolute Distance from the Mean}

In the model above, an agent's distinctiveness preference was defined purely in terms of the population's dispersion (specifically, in units of standard deviation). We found that a consequence of this postulate is that the population variance either approaches zero (global conformity) or diverges indefinitely. In real populations, a stable level of diversity may be maintained (e.g., a stable political spectrum) without either complete conformity or ever-widening divergence. What is the simplest and most natural way to endow the model with this capacity? 

In the preceding variant, where a multiple of standard deviation was the only distinctiveness metric, we saw that agents can converge to the identical position -- which perforce {\em is} the mean. They have no problem with being average in that case. We will see that if we add to the previous framework a little repulsion from the mean, the dynamics are altered substantially. In particular, diversity can be maintained.

We need only revise our formula for an agent's ideal position as follows:
\begin{equation}
x_i^* (t) = \bar{x}(t) + \delta_i [\sigma(t) + \epsilon],
\end{equation}
where $\epsilon$ is a positive constant. Thus, even if the population were at global conformity, an agent would prefer to be $\delta_i \epsilon$ units away from the mean. Note that as long as $\epsilon$ is small, distinctiveness is still approximated in terms of standard deviation when the variance is large. However, when the variance is small, the new mean-repelling term will dominate. This model is otherwise identical to Model 1 (i.e.,  Model 1 is a special case of Model 2). How does this small revision change the dynamics? 

If all individuals have the same distinctiveness preferences (i.e., $\forall i, \delta_i = \delta$), we get conformity as before. However, instead of stabilizing at a single position, the population continues to move, as persistent feedback to be $\delta_i \epsilon$ units from the mean pushes the population ever upwards (Fig. 5A) (or downwards, if $\delta < 0$). Things become more interesting when preferences for distinctiveness are heterogeneous. For several cases of two values of $\delta$, it can be shown analytically that the distance between the two groups of agents will stabilize at a nonzero value (see Appendices D and E). 

\subsection{Traveling Waves}

Simulations further show that diversity will be stably maintained in populations with a wider range of distinctiveness preferences, where stability is defined as the absence of consistent change in the population positional variance after an initial transient period of reorganization (in this transient period, agents effectively ``sort" by $\delta$ order). The absence of convergence is not the same thing as the absence of movement, however. Small asymmetries in the distributions of distinctiveness preferences and/or initial positions can lead to {\em stable relative positions but a continuously changing population mean}, as in the traveling wave depicted in Fig. 5B. 

\begin{figure*}[tp]
\includegraphics[width=0.99\textwidth]{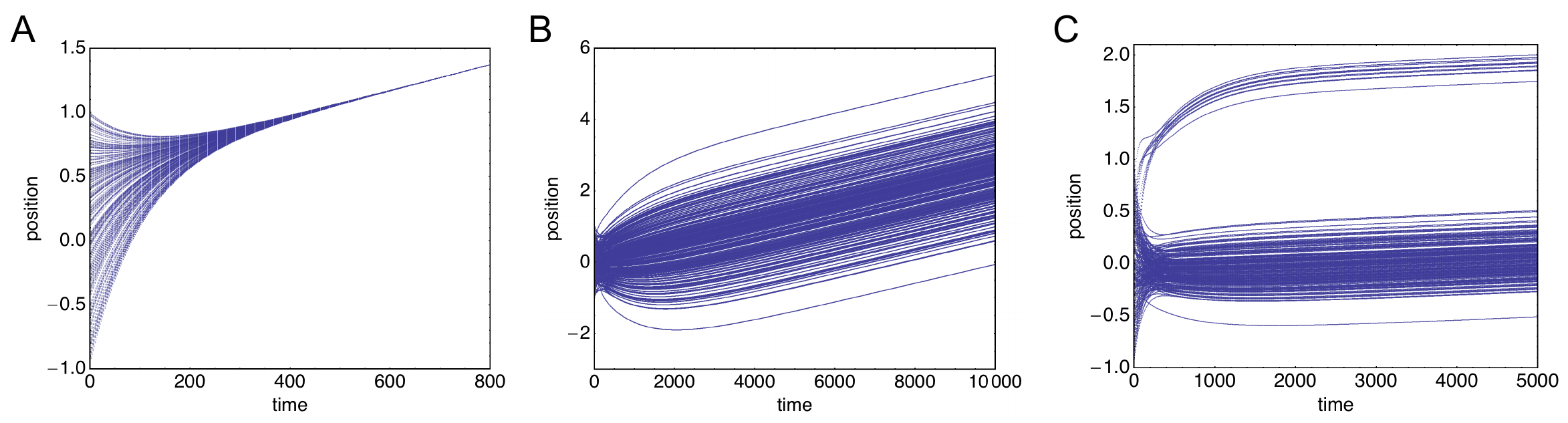}
\caption{\label{fig:fig5} Example trajectory plots for Model 2, patterned after otherwise identical runs for Model 1. (A) Uniform distinctiveness preferences, $\delta = 1$. (B) Normally distributed preferences drawn from $N$(0, 0.9). (C) Bimodally distributed preferences. 92\% conformists' preferences drawn from $N$(Ð0.2, 0.3). 8\% nonconformists' preferences drawn from $N$(3, 0.1). For all runs, $\epsilon = 0.1$. }
\end{figure*}

We also found that our model extension could stabilize the clusters of conformists and nonconformists described in the previous section. Fig. 5C depicts a model run under conditions identical to Fig. 4B, but with the new model extension. Instead of converging to global conformity as before, positional heterogeneity is stable in the population, resulting in persistent majority and minority ``factions."

\section{\label{sec:level1}Discussion}

This paper has attempted to bring increased rigor and explicit modeling to the general field of individual distinctiveness and social dynamics. Using very simple notions of distinctiveness preferences and simple rules of adjustment, we developed two models, each of which has produced several new results (see Table A.1 for a summary). 

Model 1 demonstrates an unexpected generative mechanism for a ubiquitous and important social phenomenon: conformity. Individual adaptation to {\em increase} inter-agent similarity is, of course, sufficient to generate conformity \cite{asch51, degroot74, bern94, hen98, cial04, dand13}.  But we demonstrate that it is not necessary, and indeed show that {\em the quest for distinctiveness can also generate conformity}. Conformity of course occurs when people strive for similarity, but it evidently can also occur when people strive for distinctiveness. Perhaps, then, we should not be surprised at its ubiquity.

Moreover, this conformity was not dependent on our initial modeling assumption of identical distinctiveness preferences. We showed that as long as the variance in individuals' preferences is not too large, the population still converges toward conformity even with heterogeneous distinctiveness preferences. We then explored bimodally distributed distinctiveness preferences, finding that a small minority of nonconformists (fewer than 10\%) can significantly change the position even of a large conformist majority. As in human history, so in the model: extremists can matter. Of course, society also exhibits stable diversity.

Model 2 generalized Model 1 to include (absolute) repulsion from the mean, providing a simple mechanism sufficient to generate and maintain diversity. Unexpectedly, the model also produces travelling waves in which the positional distribution retains its form while moving to the right over time.  For bimodally distributed distinctiveness preferences, clusters akin to ``factions" emerge and are sustained. 

All in all, a very simple explicit model was shown to produce a wide range of unexpected results, the central one being that conformity can emerge despite individual preferences for distinctiveness -- indeed, because of them!  Further extensions such as the addition of space, agent movement, networks, and multiple dimensions (beyond our single positional one) would doubtless enrich the dynamics and repay study, as would the addition of noise or bias to the agents' assessments of positional distributions. Contrary to inductivist legend, it often occurs in the history of science that theoretical work {\em precedes} and guides empirical activity \cite{epstein08}.  We would, of course, be delighted were the present theoretical work to have the same effect. Meanwhile, this elementary model should contribute to the literature on social dynamics, by providing ($i$) mathematically specific definitions of individual distinctiveness, ($ii$) simple agent adaptations meant to attain them, and ($iii$) the collective dynamics that result.

\section*{Acknowledgments}
\noindent For close mathematical scrutiny, we thank Jon Parker, Julia Chelen, and Gerard Weisbuch. For valuable comments and stimulating discussions, we also thank Jimmy Calanchini, Brett Calcott, Peter Duggins, Erez Hatna, Matthew Jarman, Eili Klein, and Michael Makowsky. PES and JME conceived of the study, created and analyzed the mathematical model, and wrote the manuscript. PES derived mathematical proofs, performed computational analyses, and made the figures. This work was supported by NIH Pioneer Award DP1OD003874 to Joshua M. Epstein.

\bibliographystyle{rspublicnatwithsort} 
\bibliography{distinctlyconformist}{}

\appendix
\setcounter{figure}{0} \renewcommand{\thefigure}{A.\arabic{figure}}
\setcounter{table}{0} \renewcommand{\thetable}{A.\arabic{table}}

\section{The Model}

An agent's ideal location is: 
\begin{equation} \label{eq:model}
x^*_i (t) = \bar{x}(t) + \delta_i \left[ \sigma(t) + \epsilon \right],
\end{equation}
where $\bar{x}(t)$ is the mean location of the agents in the population, $\sigma(t)$ is the standard deviation of those locations, $\delta_i$ is the distinctiveness preference of agent $i$, and $\epsilon$ is a factor representing absolute repulsion from the population mean. Model 1 is defined by $\epsilon = 0$, and Model 2 by $\epsilon > 0$.

At each time step, the agent updates its position according to
\begin{equation} \label{eq:update}
\Delta x_i = k \left[ x^*_i (t) - x_i (t) \right],
\end{equation}
where $x_i(t)$ is the current position of agent $i$ and $k \in (0,1]$ is an adjustment rate. 

In the main text, Figure 1C illustrates the point of convergence for simulations in which all agents have the same distinctiveness preferences ($\forall i, \delta_i = \delta$) for several different initial distributions of agent positions. Agents' initial positions were initially randomly drawn from a modified Beta distribution with support $[-1, 1]$. This transformation was necessary, because in the standard Beta distribution, values are instead drawn from $[0, 1]$. Figure S1 illustrates the shape of these distributions. For all other simulations discussed in the main text, initial positions were drawn from a uniform distribution in $[-1, 1]$. 

\begin{figure}
\includegraphics[width=0.69\textwidth]{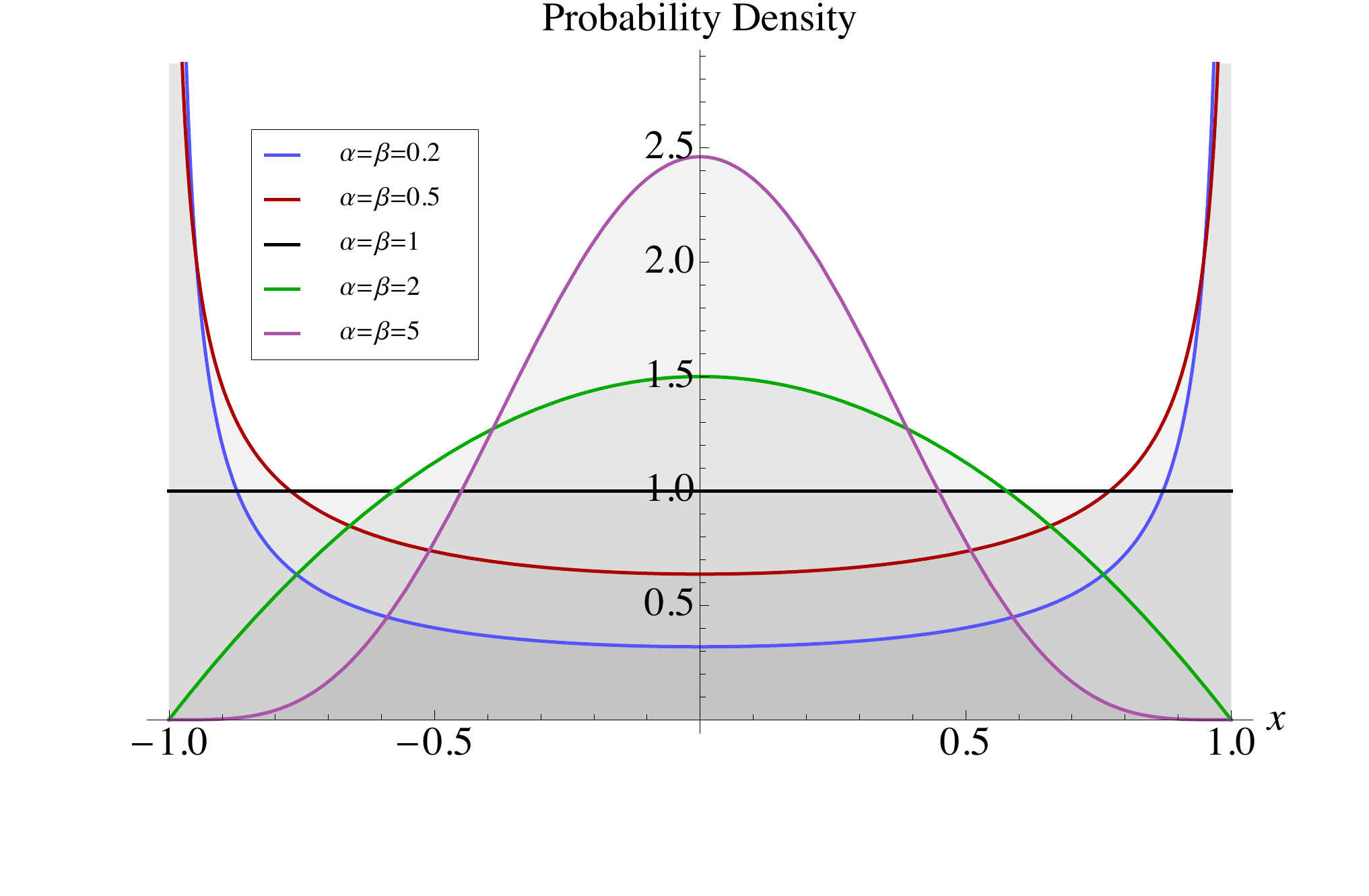}
\centering
\caption{
Probability densities of initial distributions of agent positions featured in the main text, Fig. 1C. Distributions are Beta distributions, transformed so that the support is $[-1, 1]$. 
}
\label{spatial}
\end{figure}

\section{Derivation of the change in population mean}
Here we derive an equation for the change in the population mean for Model 1. Most generally, we start with 
\begin{equation}\label{eq:deltamean}
\Delta \bar{x} = \bar{x}(t + 1) - \bar{x}(t).
\end{equation}
From equations \ref{eq:model} and \ref{eq:update}, it follows that for an agent $i$, 
\begin{equation} \label{eq:singleupdate}
x_i (t + 1) = x_i (t) + k \left[ \bar{x}(t) + \delta_i \sigma(t) - x_i (t) \right].
\end{equation}
Summing over all agents gives us
\begin{align} \label{eq:meant}
\bar{x}(t + 1) 	&=	\frac{1}{N} \sum_i x_i (t+1) \nonumber \\
					&= \frac{1}{N} \sum_i x_i (t) + \frac{k}{N} \sum_i \left[  \bar{x}(t) + \delta_i \sigma(t) - x_i (t) \right] \nonumber \\
					&= \bar{x}(t) + \frac{k}{N} \sum_i \left[  \bar{x}(t) + \delta_i \sigma(t) - x_i (t) \right]. 
\end{align}
Then, by equations \ref{eq:deltamean} and \ref{eq:meant}, we obtain the result shown in the main text, 
\begin{align}
\Delta \bar{x} 	&= \frac{k}{N} \sum_i \left[  \bar{x}(t) + \delta_i \sigma(t) - x_i (t) \right] \nonumber \\
					&= \frac{k \sigma(t)}{N} \sum_i  \delta_i .
\end{align}

\section{Proof that the population always converges when all agents have the same distinctiveness preference}

Under Model 1, let all agents have the same distinctiveness preference, i.e., $\forall i, \delta_i = \delta$. Since each individual records the same population mean and standard deviation, it follows that at time $t$, all agents will have the same ideal position: 
\begin{equation} \label{eq:allDeltaUpdate}
x^*(t) = \bar{x}(t) + \delta \sigma (t).
\end{equation}
By equations \ref{eq:update} and \ref{eq:allDeltaUpdate}, the update rule for each agent is given by
\begin{equation}
x_i (t + 1) = x_i (t) + k \left[ \bar{x}(t) + \delta \sigma (t) - x_i (t) \right].
\end{equation}
We will prove that this rule leads the population to converge by showing that the variance of agents' positions at time $t + 1$ is always less than the variance of agents' positions at time $t$. Recall that the equation for variance is
\begin{equation}\label{eq:variance}
\sigma^2 = E[(x^2)] - \bar{x}^2. 
\end{equation}
At time $t+1$, the variance is given by
\begin{equation} \label{eq:variancetplus}
\sigma^2 (t + 1) =   \frac{1}{N} \sum_i \underbrace{x_i^2 (t+1)}_{P} - \underbrace{\bar{x}^2(t+1)}_{Q}
\end{equation}

We decompose the problem by first solving for $P$ and $Q$, as shown in equation \ref{eq:variancetplus}. To make the derivation cleaner, we'll drop the ``$(t)$" when indicating variables taken at time $t$. We start by solving for $Q$.
\begin{align}
\bar{x}^2(t+1) 	&= \left[ \frac{1}{N} \sum_i \left( x_i + k[ \bar{x} + \delta \sigma - x_i] \right) \right]^2 \\
					&= \left[ \bar{x} + k[\bar{x} + \delta \sigma - \bar{x}] \right]^2 \\
					&= \left[ \bar{x} + k \delta \sigma \right]^2 \\
					&= \bar{x}^2 + k^2 \delta^2 \sigma^2 + 2k\delta \sigma \bar{x}
\end{align}

Now we solve for $P$.
\begin{align}
x_i^2 (t+1)	&= \left[ x_i + k \bar{x} + k \delta \sigma - kx_i \right]^2 \\
				&=  x_i^2 + 2k\bar{x}x_i + 2k \delta \sigma x_i - 2kx_i^2 \nonumber \\	
				& \qquad {} + k^2 \bar{x}^2 + 2k^2 \delta \sigma \bar{x} - 2k^2 \bar{x}x_i  \nonumber \\
				& \qquad {} +  k^2 \delta^2 \sigma^2 - 2k^2 \delta \sigma x_i + k^2 x_i^2  \nonumber \\
				&= (1-k)^2 x_i^2 + k^2 \delta^2 \sigma^2 + k^2 \bar{x}^2 + 2k[\bar{x}x_i + \delta \sigma x_i]   \nonumber \\
				& \qquad {} + 2k^2[ \delta \sigma \bar{x} - \bar{x}x_i - \delta \sigma x_i]
\end{align}

Taking the expectation of $P$ (summing over all the $i$'s and dividing by $N$) gives us:
\begin{equation}
E[(x^2(t+1))] = (1-k)^2 E[(x^2)] + k^2 \delta^2 \sigma^2 + 2k \bar{x}^2 + 2k \delta \sigma \bar{x} - k^2 \bar{x}^2
\end{equation}

Reassembling these components, we see that 
\begin{align} \label{eq:newVar}
\sigma^2 (t + 1)	&= 	E[(x^2(t+1))] - \bar{x}^2(t+1) \nonumber \\
						&= 	(1-k)^2 E[(x^2)] + k^2 \delta^2 \sigma^2 + 2k \bar{x}^2 + 2k \delta \sigma \bar{x} - k^2 \bar{x}^2  \nonumber \\
						& \qquad {} - \bar{x}^2 - k^2 \delta^2 \sigma^2 - 2k\delta \sigma \bar{x} \nonumber \\
						&=  (1-k)^2 E[(x^2)]  - (1-k)^2 \bar{x}^2 \nonumber \\
						&= (1-k)^2 \left[ E[(x^2)]  - \bar{x}^2 \right].
\end{align}

Combining equations \ref{eq:variance} and \ref{eq:newVar} produces the equation given in the main text:
\begin{equation} \label{eq:varresult}
\sigma^2 (t + 1) = (1-k)^2 \sigma^2 (t).
\end{equation}
Since $k$ is bounded in (0, 1], equation \ref{eq:varresult} implies at each time step, the variance of agent positions will  be strictly less than  in the previous time step. In other words, the population variance will always converge to zero: agents will occupy the same position, as claimed.

\section{Proof that up to a critical value of $\delta_N$, when there are conformists and nonconformists, positions  converge }

Assume two homogenous subpopulations: {\em conformists} (C) and {\em nonconformists} (N). Conformists want to be like the average individual, so $\delta_C = 0$. Nonconformists want to be different (but to an identical extent), so $\delta_N = \delta > 0$. The frequency of conformists in the population is $p$, so the frequency of nonconformists is $1 - p$. We will work out the math using the assumptions of Model 2, but  will show that the results specialize to Model 1 (by setting $\epsilon = 0$). 

The mean position is
\begin{equation}
\bar{x} = p x_C + (1 - p) x_N,
\end{equation}
and the variance of positions is
\begin{equation}
\sigma^2 = p (1-p) (x_N - x_C)^2 ,
\end{equation}
so the standard deviation is
\begin{equation}
\sigma = \left[ p(1-p) \right]^ \frac{1}{2} (x_N - x_C).
\end{equation}

Now we solve for how the agents' positions change over time. 

\begin{align}
\Delta x_C 	&= k \left[ \bar{x} - x_C  \right]\\
				&= k \left[ p x_C + (1-p) X_N - x_C \right]\\
				&= k (1-p)(x_N - x_C)
\end{align}

\begin{align}
\Delta x_N 	&= k \left[ \bar{x} - x_N + \delta (\sigma + \epsilon)  \right]\\
				&= k \left[ p x_C + (1-p) x_N - x_N + \delta [p(1-p)]^\frac{1}{2} (x_N - x_C) + \delta \epsilon  \right]\\
				&= k \left[ \left(-p +  \delta [p(1-p)]^\frac{1}{2} \right) (x_N - x_C) + \delta \epsilon \right]
\end{align}

We are interested in convergence and divergence of the two populations, so let us define a variable $D$ to be the distance between conformists and nonconformists, and observe how this distance changes over time. 
\begin{align} \label{eq:deltaD}
\Delta D 	&= \Delta x_N - \Delta x_C \nonumber \\
			&= k \left[ \left( \delta [p(1-p)]^\frac{1}{2} - 1 \right) (x_N - x_C) + \delta \epsilon \right]
\end{align} 
An equilibrium will exist when $\Delta D = 0$. Setting equation \ref{eq:deltaD} to zero yields  
\begin{equation} \label{eq:conformists}
x_N - x_C = \frac{\delta \epsilon}{1 - \delta [p(1-p)]^\frac{1}{2}}
\end{equation}

When $\epsilon > 0$ (Model 2), we can get stable coexistence of two  populations separated by a fixed distance. If $\epsilon = 0$ (Model 1), then the population will either converge to a single value, or diverge.  We began by assuming that $x_N > x_C$, and there is no mechanism by which this relationship can reverse. Therefore, stability/convergence will only occur if the denominator of the right-hand side of equation \ref{eq:conformists} is positive. More precisely, the condition for convergence in Model 1 and stability in Model 2 is
\begin{equation}
\delta < \frac{1}{\sqrt{p(1-p)}}.
\end{equation}
This implies the separatrix equation (Eq. 8) of the main text. Importantly, when $\epsilon > 0$ (Model 2), stability does not imply that the agents remain in the {\em same} positions over time, only that they are in the same {\em relative} positions -- that is, the distance between the conformist and nonconformist populations does not change. Indeed, the population can get caught in a feedback loop in which the average position increases indefinitely, as illustrated in the main text.

\section{Stability and convergence with two agents possessing equal but opposite distinctiveness preferences}

Consider a very simple case of Model 2: a population of two agents, each with equal but opposite distinctiveness preferences (i.e. $\delta_1 = -\delta_2$). For simplicity, let $\delta_1 = \delta$ and $\delta_2 = -\delta$, and assume $\delta \geq 0$. We will further assume that at $t = 0$, $x_1 > x_2$. Due to the relationship between the two agents' distinctiveness preferences, this ordering will not change. 
The agents' mean position is given by $\bar{x} = \frac{x_1 + x_2}{2}$, and their 
standard deviation by $\sigma = \frac{1}{2} (x_1 - x_2)$.

The agents update their positions according to the following:
\begin{align}
\Delta x_1 	&= k \left[ \bar{x} - x_1 + \delta_1 (\sigma + \epsilon) \right]\\
				&= k \left[ \frac{x_1 + x_2}{2} - x_1 + \frac{\delta}{2} (x_1 - x_2) + \delta \epsilon \right] 
\end{align}
and
\begin{align}
\Delta x_2 	&= k \left[ \frac{x_1 + x_2}{2} - x_2 - \frac{\delta}{2} (x_1 - x_2) - \delta \epsilon \right]. 
\end{align}

Let us define $D$ as the distance between the two agents, so that $D = x_1 - x_2$. 
This distance changes according to: 
\begin{align} \label{eq:deltaD2}
\Delta D 	&= \Delta x_1 - \Delta x_2 \nonumber \\
			&= k \left[ x_2 - x_1 + \delta(x_1 - x_2) + 2 \delta \epsilon  \right] \nonumber \\
			&= k \left[ (\delta - 1)(x_1 - x_2) + 2 \delta \epsilon  \right]
\end{align}

We want to know if and where the equilibrium lies. In other words, when do the agents stop moving relative to one another, so that $\Delta D = 0$? Setting the right-hand side of equation \ref{eq:deltaD2} to zero, we obtain
\begin{equation} \label{eq:ebo}
x_1 - x_2 = \frac{2 \delta \epsilon}{1 - \delta}.
\end{equation}

Equation \ref{eq:ebo} implies two things. First, if $\epsilon = 0$ (Model 1), then the only equilibrium occurs when the agents occupy the same location. If the distance between the two agents is growing, it will continue to increase. Second, if $\epsilon > 0$ (Model 2), then the agents will reach a stable relative positioning if and only if $\delta < 1$. 

It is noteworthy that if $\delta = 1$ (a singularity of equation \ref{eq:ebo}), then the standard deviation of the $\delta$s is also 1, which was the transition point between convergence and divergence for normally distributed distinctiveness preferences obtained via simulations in the main text.    

\clearpage
\begin{table*}
\begin{tabular}{ | l |  p{9cm}| }
\hline
\multicolumn{2}{ |l| }{MODEL 1: DISTINCT. PREFERENCES IN UNITS OF STANDARD DEVIATION} \\
\hline
{\bf Condition} & {\bf Result} \\ \hline
All agents have the same $\delta$ & When all agents have the same $\delta$,  positions converge to a single point, approaching global conformity despite a fixed preference for distinctiveness \\ \hline
\multirow{2}{*}{Heterogeneous $\delta$s (general)} & Agents converge to a single position, as long as the variance in distinctiveness preferences is not too large. When the variance of $\delta$s is large enough, the population diverges.
\\ \cline{2-2}
&  An unstable equilibrium also exists in which all agents are initialized in their ideal positions.  Minor perturbations will lead to either convergence or divergence. \\ \hline
Normally distributed $\delta$s & For normally distributed $\delta$s with a mean of zero, bifurcation between convergence and divergence occurs at a SD of 1. Near this value there is high sensitivity to initial conditions regarding convergence or divergence. For convergent runs, the mean position can move a large distance from its initial value before the population converges. \\ \hline
Bimodally distributed $\delta$s & A relatively small minority of nonconformists with large $\delta$ can significantly change the position of an overwhelming majority of agents who have very small delta (conformists).\\ \hline
All convergent runs & The limit point of convergence depends on both $\delta$ and the initial distribution of individuals' positions.
 \\ \hline \hline
\multicolumn{2}{ |l| }{MODEL 2: DISTINCT. PREFS. INCLUDE ABSOLUTE DISTANCE FROM MEAN} \\ \hline
{\bf Condition} & {\bf Result} \\ \hline
All agents have the same $\delta$ & Agents converge to the same position, but this position continues to move.  \\ \hline
\multirow{2}{*}{Heterogeneous $\delta$s} & Diversity is maintained, as the standard deviation converges to a nonzero constant.
\\ \cline{2-2}
& The population mean may continue to change, resulting in a ``traveling wave" of positions. \\ \hline
Bimodally distributed $\delta$s & Diversity can be maintained with persistent ``factions." \\ \hline
\end{tabular}
\caption{Summary of results for Models 1 and 2}
\end{table*}

\end{document}